\documentclass[journal,twoside,web]{ieeecolor}
\usepackage{generic}
\usepackage{multirow}
\usepackage{amsmath,amssymb,amsfonts}
\usepackage{algorithmic}
\usepackage{graphicx}
\usepackage{algorithm,algorithmic}
\usepackage[hidelinks]{hyperref}
\usepackage[sorting=none]{biblatex}
\usepackage{textcomp}
\def\BibTeX{{\rm B\kern-.05em{\sc i\kern-.025em b}\kern-.08em
    T\kern-.1667em\lower.7ex\hbox{E}\kern-.125emX}}
\usepackage{booktabs} 
\usepackage{float}    
\usepackage{caption}  
\usepackage{booktabs} 
\usepackage{xcolor}
\newcommand{\differencecolor}{black}  

\begin{document}
\title{EEG2GAIT: A Hierarchical Graph Convolutional Network for EEG-based Gait Decoding}
\author{Xi Fu, \IEEEmembership{Student Member, IEEE}, Rui Liu, Aung Aung Phyo Wai, Hannah Pulferer, Neethu Robinson, Gernot R Müller-Putz, Cuntai Guan, \IEEEmembership{Fellow, IEEE}
\thanks{This work was supported by the RIE2020 AME Programmatic Fund, Singapore (No. A20G8b0102). \textit{(Corresponding author: Cuntai Guan.)}
}
\thanks{Xi Fu, Rui Liu, Aung Aung Phyo Wai, Neethu Robinson are with the College of Computing and Data Science, Nanyang Technological University, Singapore 639798 E-Mail: fuxi0010@e.ntu.edu.sg; \{rui.liu, apwaung, nrobinson, ctguan\}@ntu.edu.sg}
\thanks{Hannah Pulferer and Gernot R Müller-Putz are with the Institute of Neural Engineering, Graz University of Technology, Graz, Austria. E-Mail: \{hannah.pulferer, gernot.mueller\}@tugraz.at}
\thanks{Cuntai Guan is with the College of Computing and Data Science, and also the Center for Al in Medicine (C-AlM) at Nanyang Technological University, Singapore.}
\thanks{All machine learning codes used in this study have been deposited in the GitHub repository and can be found with the following link: https://github.com/FuXi1999/EEG2GAIT.git
}
}

\maketitle
\begin{abstract}
Decoding gait dynamics from EEG signals presents significant challenges due to the complex spatial dependencies of motor processes, the need for accurate temporal and spectral feature extraction, and the scarcity of high-quality gait EEG datasets. To address these issues, we propose EEG2GAIT, a novel hierarchical graph-based model that captures multi-level spatial embeddings of EEG channels using a Hierarchical Graph Convolutional Network (GCN) Pyramid. To further improve decoding \textcolor{\differencecolor}{ performance}, we introduce a Hybrid Temporal-Spectral Reward (HTSR) loss function, which \textcolor{\differencecolor}{integrates} time-domain, frequency-domain, and reward-based loss components. \textcolor{\differencecolor}{In addition}, we contribute a new Gait-EEG Dataset (GED), consisting of synchronized EEG and lower-limb joint angle data collected from 50 participants \textcolor{\differencecolor}{across two laboratory visits}. \textcolor{\differencecolor}{Extensive experiments demonstrate that EEG2GAIT with HTSR achieves superior performance on the GED dataset, reaching a Pearson correlation coefficient ($r$) of 0.959, a coefficient of determination ($R^2$) of 0.914, and a Mean Absolute Error (MAE) of 0.193. On the MoBI dataset, EEG2GAIT likewise consistently outperforms existing methods, achieving an $r$ of 0.779, an $R^2$ of 0.597, and an MAE of 4.384. Statistical analyses confirm that these improvements are significant compared to all prior models.} Ablation studies \textcolor{\differencecolor}{further} validate the contributions of the hierarchical GCN modules and \textcolor{\differencecolor}{the proposed} HTSR loss, while \textcolor{\differencecolor}{saliency analysis highlights the involvement of} motor-related brain regions in decoding tasks. \textcolor{\differencecolor}{Collectively, these} findings underscore EEG2GAIT's potential for advancing brain-computer interface applications, particularly in lower-limb rehabilitation and assistive technologies.

\end{abstract}

\begin{IEEEkeywords}
Deep Learning, electroencephalographyy, graph neural networks, movement decoding, lower limb movement
\end{IEEEkeywords}

\section{Introduction}
\label{sec:introduction}
Gait is a fundamental component of human motor behavior, relying on not only motor control systems but also cognitive and sensory processes~\cite{schmidt2018motor}. The electroencephalogram (EEG), which records electrical activity from the scalp, provides a non-invasive way to monitor these neural dynamics in real time. Thus, decoding gait patterns from EEG signals is a critical step toward understanding the complex interplay between brain activity and motor functions, offering valuable insights into the neural mechanisms underlying movement~\cite{pfurtscheller1977event}. This capability is particularly crucial in neural rehabilitation, where accurate decoding of movement-related brain signals enables patients to control external devices with greater precision~\cite{wang2012self}. Moreover, accurate gait decoding facilitates the development of new therapeutic approaches for various neurological conditions, including stroke~\cite{baniqued2021brain}, spinal cord injury~\cite{cui2022bci}, and Parkinson’s disease~\cite{romero2024clinical}.

To achieve effective EEG-based motor decoding, various machine learning algorithms have been developed to date~\cite{ang2008filter, higashi2013common, costa2016decoding, miao2017spatial, riquelme2020better}. However, the mentioned works largely focused on manual feature extraction from the EEG, failing to extract highly discriminative features unique to each participant and thus limiting the decoding performance. Deep learning algorithms have addressed these limitations to some extent by extracting subject-specific EEG patterns in a data-driven manner, thereby enhancing the decoding performance. For instance, long short-term memory (LSTM) networks~\cite{wang2018lstm} have been utilized to capture the temporal dependencies in EEG signals, reflecting the dynamic nature of brain activity during gait cycles. Convolutional neural networks (CNNs) have likewise been employed to automatically learn spatial-temporal features from raw EEG signals, improving the identification of movement-related patterns~\cite{schirrmeister2017deep, lawhern2018eegnet}. These advanced models leverage the intricate relationship between neural activations and motor execution, enabling more precise and robust decoding of gait patterns.

Building upon this foundation, neurokinematics research has further highlighted the coordinated activation of different brain regions during various stages of movement, each contributing to distinct cognitive processes~\cite{reddy2018brain}. Numerous studies have explored brain region activations during motor tasks using EEG, providing valuable insights into the underlying neurophysiological processes. Event-related desynchronization (ERD) and synchronization (ERS) have been widely used to characterize motor cortex activity during movement planning and execution~\cite{pfurtscheller1999event}. For instance, $\mu$ and $\beta$ rhythms, which are associated with motor actions, show distinctive patterns of modulation during motor tasks, highlighting the dynamic activation of cortical regions~\cite{neuper2001event}. Additionally, EEG changes observed during motor skill acquisition, particularly in the $\alpha$ and $\beta$ bands, reflect cortical plasticity and the reorganization of motor areas as learning progresses, a process influenced by both age and task complexity~\cite{bootsma2021neural}. The negative correlation between EEG $\alpha$/$\beta$-band activity and BOLD signals in sensorimotor areas during motor tasks has been shown to provide valuable insights into the dynamics of motor cortex engagement~\cite{yuan2010negative}. The use of EEG to reconstruct three-dimensional hand movements has further demonstrated its efficacy in real-time decoding of motor region activity~\cite{bradberry2010reconstructing}. Inspired by these findings, several deep learning approaches have incorporated prior neurophysiologal knowledge to improve motor decoding performance. Functional topological relationships between electrodes are leveraged to decode time-resolved EEG motor imagery signals using Graph Convolutional Networks (GCNs), as demonstrated in prior work~\cite{hou2022gcns}. Local and global biological topology among brain regions has also been utilized for EEG-based emotion recognition through graph neural networks with regularization, as shown by Zhong and colleagues~\cite{zhong2020eeg}. These graph-based methods highlight the benefits of treating EEG electrodes as a non-Euclidean structure, enabling enhanced motor decoding performance. 

Despite these advances, however, additional neurophysiological priors remain underexplored, particularly in the context of gait pattern decoding. Particularly the existing loss functions may be suboptimal for accurate gait prediction. Traditional regression models often overlook the temporal dependencies between separate time steps inherent in continuous gait, assuming each prediction as independent instead. This considerably limits their capacity to capture the full context of neural dynamics. Furthermore, current deep learning regression models tend to prioritize samples with large rather than small target deviations. This behavior is driven by the characteristics of the often-used Mean Square Error (MSE) loss, which generates smaller gradients for well-predicted samples and thus results in smaller updates during training. However, in EEG-based gait prediction, every time segment can contain crucial information. As a result, ignoring well-predicted samples may lead to the underutilization of important global features. Moreover, EEG signals are highly sensitive to subtle physiological changes, including minor variations in gait patterns. These nuanced fluctuations critically influence accurate analysis and interpretation of the data. However, traditional loss functions, such as MSE or Mean Absolute Error (MAE) losses, are typically designed to minimize overall prediction errors and may not effectively capture or emphasize these fine-grained nuances. Consequently, traditional loss functions often fail to capture intricate, low-amplitude variations in EEG signals that are crucial for understanding the underlying physiological changes. In turn, the model's ability to accurately identify these subtle features are limited, affecting both its prediction accuracy and generalization performance.  
All of these mentioned limitations emphasize the need for novel algorithms that can fully leverage both the spatial and temporal dynamics of EEG data to achieve more accurate and robust gait pattern decoding.

To address the shortcomings of existing models, we propose the new model EEG2GAIT which introduces a Hierarchical GCN Pyramid (HGP) for EEG-based gait regression. This model is specifically designed to capture the intricate spatial relationships among EEG channels by extracting multi-level spatial embeddings that effectively represent the dynamic brain activity involved in gait. 

To further improve decoding accuracy, we propose a novel loss function, the Hybrid Temporal-Spectral Reward (HTSR) Loss, specifically designed for time series modelling. This loss function integrates three essential components: an MSE Loss, a Time-Frequency Loss, and a Reward Loss. The MSE Loss serves as the core metric, ensuring that the predicted gait values align closely with the actual data in the time domain. The time-frequency component enhances this by capturing both temporal and spectral characteristics of gait patterns, enabling the model to account for long-term dependencies and dynamic patterns often overlooked by conventional time-series models. Additionally, the Reward Loss encourages the model to focus on well-predicted samples, preventing underutilization of global features and ensuring balanced learning from both large and small errors. By addressing subtle variations in gait patterns, HTSR Loss aims to improve model generalizability across different datasets, offering a robust and comprehensive approach for continuous gait decoding from EEG signals.

To enrich the existing body of high-quality gait datasets, we also contribute a new walking dataset comprising EEG and joint angle data from 50 able-bodied participants. The dataset includes natural gait data collected from each participant across two visits to the laboratory on different days.
\begin{color}{\differencecolor}
    
\textcolor{\differencecolor}{In this work, we propose EEG2GAIT, a novel framework designed to decode gait patterns from EEG signals.} \textcolor{\differencecolor}{We benchmark EEG2GAIT against several state-of-the-art baselines and conduct comprehensive ablation studies to analyze the contributions of individual modules.} \textcolor{\differencecolor}{Additionally, we incorporate saliency map visualization to investigate the spatial characteristics of learned neural representations and to better understand the relationship between EEG features and motor behavior.}

Main contributions of this paper are summarized as follows:
\begin{itemize}
    \item \textbf{\textcolor{\differencecolor}{EEG2GAIT Model:}} \textcolor{\differencecolor}{A novel model incorporating a Hierarchical Graph Convolutional Network Pyramid for EEG-based gait regression, effectively capturing complex spatial relationships in EEG data.}
    \item \textbf{\textcolor{\differencecolor}{Hybrid Temporal-Spectral Reward (HTSR) Loss:}} \textcolor{\differencecolor}{A newly designed loss function for time series modelling that combines time-, time-frequency-, and reward-based objectives to enhance decoding accuracy and robustness.}
    \item \textbf{\textcolor{\differencecolor}{New Walking Dataset:}} \textcolor{\differencecolor}{A large-scale dataset collected from 50 participants over two lab visits, supporting extensive evaluation and future research in EEG-based gait decoding.}
\end{itemize}

\end{color}
\section{Related Work}
\subsection{Neural Decoding of Motor Execution Using EEG Signals}

EEG-based neural decoding of lower-limb motor execution has been widely studied, leveraging EEG's high temporal resolution for capturing motor control dynamics. Pfurtscheller and colleagues~\cite{pfurtscheller2001motor} demonstrated that event-related desynchronization (ERD) and synchronization (ERS) effectively characterize motor execution and imagery, laying the groundwork for decoding lower-limb movements. Ang and colleagues previously~\cite{ang2012filter} used common spatial pattern (CSP) and linear discriminant analysis (LDA) to decode gait phases, enhancing motor-related EEG feature extraction. Schirrmeister and colleagues~\cite{schirrmeister2017deep} employed convolutional neural networks (CNNs), boosting accuracy in classifying lower-limb motor imagery. Recent deep learning methods have improved feature extraction and spatial representation of EEG signals~\cite{goh2018spatio, tortora2020deep, fu2022matn}, demonstrating their capability to capture complex EEG patterns. Finally, Wang and colleagues~\cite{wang2018implementation} developed a real-time BCI system for lower-limb exoskeleton control, demonstrating EEG’s feasibility in rehabilitation. These studies have contributed to the development of BCIs for rehabilitation and assistive technologies.

\subsection{Graph Neural Networks}
Graph Neural Networks (GNNs) are designed for processing non-Euclidean data with a graph-like structure. One category of GNNs, known as spectral GNNs, often depends on costly eigendecomposition of the graph Laplacian~\cite{bruna2013spectral}. To address this limitation, several techniques use approximation methods for spectral filtering. For instance, ChebyNet~\cite{defferrard2016convolutional} leverages Chebyshev polynomials to approximate spectral filters, while Cayley polynomials are employed to compute spectral filters targeting specific frequency bands, scaling linearly for sparse graphs~\cite{levie2018cayleynets}. GCN simplifies spectral filtering through localized first-order aggregation~\cite{kipf2022semi}. Since EEG signals naturally exhibit a graph structure, some approaches apply GNNs/GCNs to capture spatial features from EEG data. Several methods~\cite{shamsi2021early, hou2022gcns, zhang2020motor} model brain activity using a single adjacency matrix, which provides a global representation but may not fully capture complex inter-regional interactions. Furthermore, some approaches~\cite{hou2022gcns, shamsi2021early, zhang2020motor} focus on spatial graph structures, often using averaged node features, which simplifies modeling but may reduce temporal specificity. 

\subsection{Enhancing Model Performance with Well-Classified Samples}
In regression tasks, MSE, MAE, and Root Mean Squared Error (RMSE) are commonly used as loss functions. However, these losses prioritize poorly classified samples, as well-classified ones offer limited new information for the model. Recent studies have shown that leveraging well-classified samples can enhance model performance. For example,~\cite{zhao2022well} introduces a loss function that prioritizes well-classified samples, thus enhancing feature extraction and classification accuracy. In DeepNoise~\cite{yang2022deepnoise}, well-classified samples were used to improve accuracy in noisy fluorescence microscopy images. Jung and colleagues~\cite{jung2022wbc} used generative models for white blood cell classification, demonstrating that high-quality samples improve model performance. Additionally, in EEG signal decoding, Kwon and colleagues~\cite{kwon2022friend} propose a strategy that effectively leverages well-classified samples by introducing carefully calculated friend and enemy noise to retain correct classification in friendly models while inducing misclassification in adversarial ones. These studies suggest that well-classified samples can enhance model robustness and generalization under challenging conditions.

\section{Methods and Materials}
\subsection{EEG2GAIT Overview}
\begin{figure*}
\centering
\includegraphics[width=0.8\linewidth]{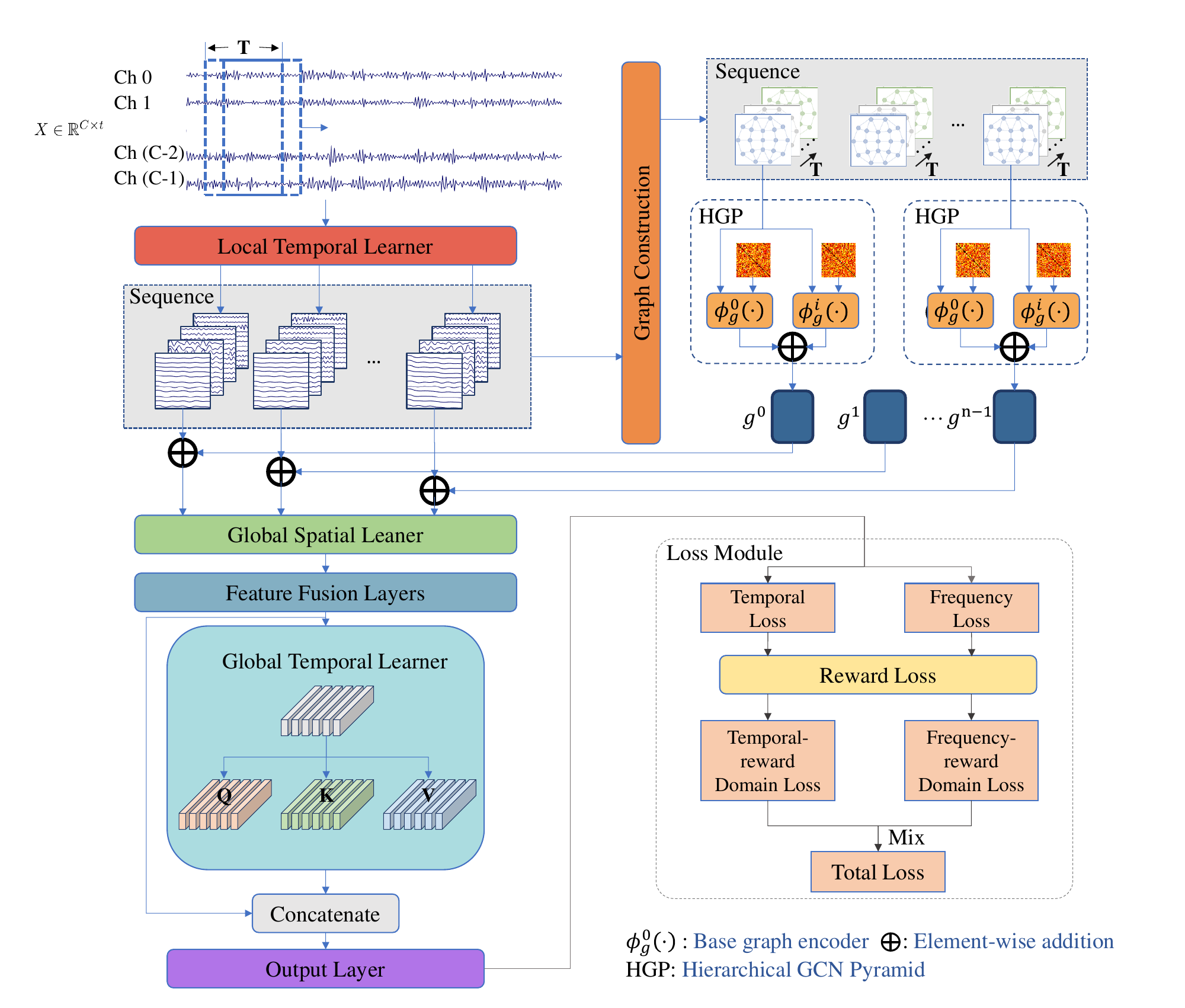}
\caption{Overview of EEG2GAIT Architecture and Loss Calculation. The sequential EEG signals are segmented into training samples \( X \in \mathbb{R}^{C \times T} \), where \( C \) is the number of channels and \( T \) is the time window length. The {Local Temporal Learner (LTL)} uses 1D convolutions along the temporal dimension to capture local temporal information. \textcolor{\differencecolor}{Each EEG sequence is divided into $n$ segments, and the LTL output is then reorganized into $n$ temporal graphs, each representing one segment.} These graphs are input to the Hierarchical GCN Pyramid ({HGP}), which transforms each graph into a token embedding. The token for all graphs and the input for graph construction will be accumulated as input to the Global Spatial Learner (GSL). GSL performs 1D convolutions across all channels. The outputs are passed into the {Feature Fusion Layers} and then Global Temporal Learner (GTL) for global temporal information extraction via self-attention. The final output goes through the {Output Layer}. Losses from both the temporal (MSE) and frequency (FFT-based L1) domains are combined via a {Reward Loss block} to yield the {Hybrid Temporal-Spectral Reward Loss}.}
\label{fig:mainfig}
\end{figure*}

In this work, we propose a novel hierarchical graph-based model, {EEG2GAIT}, specifically designed for decoding gait dynamics from EEG signals. By integrating hierarchical GCNs with temporal attention mechanisms, {EEG2GAIT} effectively captures the spatial and temporal dependencies in EEG signals. 
Figure \ref{fig:mainfig} presents the proposed architecture consisting of seven main components: Local Temporal Learner (LTL), Graph Construction Module (GCM), Hierarchical GCN Pyramid (HGP), Global Spatial Learner (GSL), Feature Fusion Layers, Global Temporal Learner (GTL), and Output Layer..

\begin{itemize}
    \item \textbf{Local Temporal Learner (LTL)}: Extracts local temporal features from EEG signals using 1D convolutional layers. This step extracts motor-related neural oscillations while reducing low-frequency noise.

    \item \textbf{Graph Construction Module (GCM)}: Transforms the outputs from LTL into temporal graph representations, where EEG channels are treated as graph nodes. The adjacency matrices for the graphs are initialized based on the spatial distances between EEG channels but are designed to be dynamically learned and updated during training. This enables the model to encode spatial relationships and adapt to the connectivity patterns of the EEG data for graph processing.

    \item \textbf{Hierarchical GCN Pyramid (HGP)}: Learns multi-level spatial embeddings from EEG graph representations through stacked GCNs. Two hierarchical graph encoders are employed to capture spatial dependencies at different levels of granularity. Each encoder uses an independent learnable adjacency matrix to dynamically model electrode connections, ensuring adaptability to different cognitive processes and brain regions. The outputs from the two encoders are fused and integrated with the original spatial-temporal features for robust spatial feature representation.

    \item \textbf{Global Spatial Learner (GSL)}: Processes fused spatial features using convolutional layers that span all EEG channels to capture global spatial dependencies. These layers use depth-wise convolution to efficiently learn frequency-specific spatial patterns while preserving the temporal structure of the data.

    \item \textbf{Feature Fusion Layers}: Combines spatial and temporal features using convolution and pooling operations. Intermediate layers refine the learned representations through dropout, normalization, and ELU activation, reducing the dimensionality and emphasizing important features for downstream processing.

    \item \textbf{Global Temporal Learner (GTL)}: Captures temporal dependencies across the entire sample time window using a multi-head self-attention mechanism. Residual connections ensure the preservation of original temporal dynamics while enhancing gradient flow during training.

    \item \textbf{Task-Specific Output Layer}: Aggregates spatial-temporal features and maps them to the predicted joint angle trajectories using a convolutional layer with constrained weights. These layers are designed to align the outputs with the specific requirements of gait decoding tasks, providing precise and robust predictions.
\end{itemize}

\subsection{Local Temporal Learner}
Let a time-series EEG signal be denoted as $(X, Y)$, where $X \in \mathbb{R}^{C \times t}$ represents the EEG signal sampled at 100 Hz, and $Y \in \mathbb{R}^{d_J \times t}$ denotes the corresponding label, with $C$ denoting the number of channels, $t$ the total number of recorded time points, and $d_J$ representing the number of lower-limb joint angles. Subsequently, a sliding window of length $T$ and step size 1 is applied to further segment the signal into shorter subsequences. As a result, $X$ is divided into $(t - T + 1)$ samples, denoted individually as $\bar{X} \in \mathbb{R}^{C \times T}$. Each $\bar{X}$ corresponds to a label $\bar{y} \in \mathbb{R}^{d_J}$.

We apply a one-dimensional convolution in the time domain on $\bar{X}$ to capture local temporal features. Specifically, we employ $F$ convolutional kernels with a size of $(1, 10)$. In this experiment, $F$ is fixed at $25$, meaning $25$ temporal filters are applied to extract meaningful temporal patterns from the EEG signal. To ensure the input and output sizes remain consistent, zero-padding is applied along the time dimension, allowing the convolution to maintain the same temporal resolution as the input.

\subsection{Graph Construction Module}

Given the output of the Local Temporal Learner, denoted as $\bar{X} \in \mathbb{R}^{F \times C \times T}$, where $F$, $C$, and $T$ represent the number of feature maps, the number of EEG channels, and the time window length respectively, the Graph Construction Module transforms $\bar{X}$ into a temporal graph representation. In this graph representation, each EEG channel is treated as a graph node, and the spatial relationships between channels are encoded as graph edges.

The adjacency matrix $A \in \mathbb{R}^{C \times C}$ defines the connectivity between the EEG channels (nodes). The initial values of $A$ are determined by retaining neighboring channels within a 30 mm radius for each EEG channel. This approach ensures that only spatially close channels are connected, reflecting the localized relationships between EEG electrodes on the scalp. The resulting adjacency matrix captures the underlying spatial structure of the EEG data effectively.

To enhance the flexibility of the graph representation, the adjacency matrix is \textcolor{\differencecolor}{implemented as a learnable parameter (nn.Parameter) during training}. \textcolor{\differencecolor}{This dynamic parameterization allows the network to adaptively refine the connectivity structure rather than relying solely on fixed priors.}

\textcolor{\differencecolor}{To provide a biologically meaningful initialization, we first construct an anatomical adjacency matrix $A_{\text{prior}} \in \mathbb{R}^{C \times C}$ based on spatial priors: two channels are connected if their inter-electrode distance is less than 30\,mm, yielding a sparse, topology-informed graph. The prior is pre-processed as:
\begin{equation}
\textcolor{\differencecolor}{\tilde{A}_{\text{prior}} = \text{ReLU}(A_{\text{prior}} + A_{\text{prior}}^\top) + I,}
\end{equation}}
\textcolor{\differencecolor}{where $I$ denotes the identity matrix (self-loops), and $\text{ReLU}$ guarantees non-negative edge weights.}

\textcolor{\differencecolor}{During training, the learnable adjacency parameter $A$ (initialized from $\tilde{A}_{\text{prior}}$) is dynamically normalized in each forward pass using symmetric normalization:}

\textcolor{\differencecolor}{
\begin{align}
D &= \text{diag}\left(\sum_j A_{ij} + \text{mask}_i\right) \\
\text{mask}_i &= \begin{cases} 
1 & \text{if } \sum_j A_{ij} = 0 \\
0 & \text{otherwise}
\end{cases} \\
\hat{A} &= D^{-\frac{1}{2}} A D^{-\frac{1}{2}}
\end{align}
}

\textcolor{\differencecolor}{where the mask ensures numerical stability for isolated nodes, and the symmetric normalization $\hat{A}$ preserves the scale of feature representations during graph convolution.}

The constructed graph representation is then passed to the HGP for multi-level spatial embedding extraction. This process enables the model to dynamically integrate both predefined and learned spatial dependencies, effectively encoding the spatial relationships among EEG channels.

\subsection{Hierarchical GCN Pyramid}
We propose a {Hierarchical GCN Pyramid (HGP)} to modulate the dynamic spatial relationships between EEG channels associated with lower limb motor processes. For each output from the {Local Temporal Learner}, the shape is $(F, C, T)$. The {Graph Construction Module} reshapes sample $\bar{X}$ into $(T, C, F)$. Each channel is regarded as one node and extracted features $f\in \mathbb{R}^F$ are regarded as node attributes. 

The HGP learns embedding $g^i$ for each subgraph $\hat{{X_i}} \in \mathbb{R}^{C \times F}$ to capture spatial features of the signal. A base graph encoder $\Phi_g(\cdot)$ is utilized to learn graph representations. In this paper, we use {GCN}~\cite{kipf2022semi} as the base graph encoder:

\begin{equation}
\Phi_g(\mathbf{H}^{(l + 1)}, \mathbf{A}) = \sigma \left( \tilde{\mathbf{D}}^{-\frac{1}{2}} \tilde{\mathbf{A}} \tilde{\mathbf{D}}^{-\frac{1}{2}} \mathbf{H}^{(l)} \mathbf{W}^{(l)} + \mathbf{b}^{(l)} \right),
\end{equation}

where:
\begin{itemize}
    \item $\mathbf{H}^{(l)}$ represents the input node features at layer $l$, with $\mathbf{H}^{(0)} = \mathbf{X}$ as the initial node features;
    \item \textcolor{\differencecolor}{$\tilde{\mathbf{A}}$ represents the symmetrically normalized adjacency matrix $\hat{\mathbf{A}}$ computed during the forward pass from the learnable adjacency parameter $\mathbf{A}$};
    \item \textcolor{\differencecolor}{$\tilde{\mathbf{D}}$ is the degree matrix used in the symmetric normalization process};
    \item $\mathbf{W}^{(l)}$ is the trainable weight matrix at layer $l$;
    \item $\mathbf{b}^{(l)}$ is the bias term at layer $l$;
    \item $\sigma(\cdot)$ is the activation function, ReLU in this work.
\end{itemize}

\textcolor{\differencecolor}{The normalization process $\tilde{\mathbf{D}}^{-\frac{1}{2}} \tilde{\mathbf{A}} \tilde{\mathbf{D}}^{-\frac{1}{2}}$ is performed dynamically during each forward pass to ensure proper scaling of the learned adjacency matrix, where isolated nodes (degree 0) are handled by adding a mask to prevent division by zero.}

To modulate the diversity of brain region connections for different cognitive processes, we propose to use multiple hierarchical GCNs, i.e., $\{\Phi^0_g(\cdot), \Phi^1_g(\cdot), \dots, \Phi^i_g(\cdot)\}$, with learnable adjacency matrices $\{A^0, A^1, \dots, A^i\}$. \textcolor{\differencecolor}{Each $A^i$ is initialized from spatially-informed priors but learns task-specific graph connections corresponding to a particular cognitive process during training.} \textcolor{\differencecolor}{The hierarchical behavior is achieved by constructing $K$ parallel GCN branches with different depths $\{L^0, L^1, \dots, L^{K-1}\}$ where $L^0 < L^1 < \dots < L^{K-1}$, while maintaining consistent hyperparameters across all branches.} For each $\Phi^i_g(\cdot)$, stacking different GCN layers allows learning different levels of node cluster similarity. Intuitively, for localized connections, such as electrodes within the same brain functional area, a deeper GCN can achieve consistent representations among these nodes. However, for global connections across different brain regions, a shallower GCN can aggregate information between these areas without oversmoothing. The deeper $\Phi^i_g(\cdot)$ becomes, the higher the level in the feature pyramid its output represents.

\subsection{Global Spatial Learner}

The output from the HGP, consisting of two graph encoders, is first integrated with the original features through a residual connection. This step combines refined spatial embeddings with the initial features to prepare the data for subsequent processing by the Global Spatial Learner (GSL).

The GSL processes these combined features using a depth-wise grouped convolution~\cite{krizhevsky2012imagenet} with a kernel size of $(C, 1)$. This operation efficiently learns spatial filters specific to each temporal filter, enabling the extraction of frequency-specific spatial patterns. To further enhance the feature representations, batch normalization~\cite{ioffe2015batch} is applied along the feature map dimension, followed by an exponential linear unit (ELU)~\cite{clevert2015fast} activation function. Dropout~\cite{srivastava2014dropout} with a probability of 0.5 is used to prevent overfitting when training on small datasets. Finally, an average pooling layer of size $(1, 3)$ reduces the temporal resolution. As a result, the output size of the GSL is \textcolor{\differencecolor}{$(F, 1, T // 3)$}.

\subsection{Feature Fusion Layers}
The Feature Fusion Layers refine spatial-temporal features through a series of convolutional blocks, consisting of three sequential convolutional blocks followed by a dropout layer. Each convolutional block applies the following operations: dropout (rate 0.5), zero-padding, convolution, batch normalization, ELU activation, and max-pooling. The kernel size and pooling size are consistent across all layers, while the number of filters varies to progressively increase feature complexity.

The first convolutional block uses 50 filters, followed by 100 and 200 filters in the second and third blocks, respectively. The kernel size is fixed at $(1, 10)$ across all blocks, with zero-padding applied to maintain the temporal dimension. Each max-pooling operation reduces the temporal resolution by a factor of $1/3$, progressively compressing the time dimension of the data.

After the three convolutional blocks, the output passes through a dropout layer with a rate of 0.5 to mitigate overfitting. The final output of the Feature Fusion Layers has a size of $(200, 1, T // 81)$, ready for subsequent temporal modeling. \textcolor{\differencecolor}{This channel dimension (200) results from the 200 kernels used in the final convolutional block.}

\subsection{Global Temporal Learner}
The Global Temporal Learner (GTL) captures temporal dynamics using a Multi-head Self-Attention (MSA) mechanism~\cite{vaswani2017attention}, which preserves the input-output shape consistency for efficient processing. Residual connections are employed to maintain the integrity of the input data and ensure stable gradient flow during training.

The input to the GTL is a tensor \( X_{\text{spatial}} \) with shape \( (200, T//81) \), representing the spatial-temporal features extracted from previous layers. To model temporal dependencies, \( X_{\text{spatial}} \) is transformed into query (\( Q_i \)), key (\( K_i \)), and value (\( V_i \)) matrices for each attention head, where \( Q_i \), \( K_i \), and \( V_i \) have dimensions \textcolor{\differencecolor}{\( (T//81, d_K) \)}. The self-attention mechanism for the \( i \)-th head is computed as:

\begin{equation}
\text{head\(_i\)} = \text{softmax}\left(\frac{Q_i \cdot K_i^T}{\sqrt{d_K}}\right) V_i 
\end{equation}

The outputs from all attention heads are concatenated along the feature dimension and passed through a linear layer to generate the final attention representation:

\begin{equation}
X_{\text{attention}} = \text{linear}(\text{Concat}(\text{head}_1, \ldots, \text{head}_h)) 
\end{equation}

Here, \textcolor{\differencecolor}{\( h \)} is the number of attention heads \textcolor{\differencecolor}{and \( d_K \) is the output dimension of each attention head.}

The self-attention mechanism evaluates pairwise correlations between the query and key representations of all time segments, enabling the model to effectively capture interdependencies across time. The output of the GTL has the same shape as its input, \( (200, T//81) \), ensuring that the temporal features are preserved while enriching the representation with inter-segment dependencies.

\subsection{Output Layer}

The Output Layer generates the final predictions by processing the features extracted by the previous modules. The output of the GTL is concatenated with the original input along the temporal dimension, resulting in a tensor of shape \( (200, T//81 \times 2) \). Each sample is then unsqueezed to add a channel dimension, forming a tensor of shape \( (200, 1, T//81 \times 2) \), which is passed to the Output Layer.

The Output Layer consists of a convolutional layer with \( d_J \) kernels, each of size \( (1, T//81 \times 2) \). This operation reduces the temporal dimension while mapping the features to the desired output space. Finally, the tensor is reshaped by removing redundant dimensions, producing the final output with a shape of \( (B, d_J) \), where $B$ and \( d_J \) corresponds to batch size and the number of joint angles predicted, respectively.

\subsection{Loss Calculation Module}

Given the model's predictions \( \hat{y}(t) \) and the ground truth \( y(t) \), the total loss is composed of a time-domain loss and a frequency-domain loss, which are weighted and combined.

\subsubsection{Time-Domain Loss}
The time-domain loss is calculated using the MSE loss:
\begin{equation}
L_{\text{time}} = \frac{1}{n} \sum_{i=1}^{n} \left( \hat{y}_i - y_i \right)^2
\end{equation}
To incentivize further learning from well-predicted samples, an additional Reward Loss term is introduced, defined as:
\begin{equation}
L_{\text{time\_reward}} = L_{\text{time}} + \beta \cdot \log\left(1 - e^{-L_{\text{time}}} + \epsilon\right)
\end{equation}
where \( \beta \) is a weighting factor for the reward term, and \( \epsilon \) is a small constant to avoid numerical instability (e.g., \( \log(0) \)).

\subsubsection{Frequency-Domain Loss}
To capture frequency-domain characteristics of the signal, the predictions and ground truth are transformed using the Discrete Fourier Transform (DFT). The frequency-domain error is computed using the \( L1 \) loss:
\begin{equation}
L_{\text{freq}} = \frac{1}{n} \sum_{i=1}^{n} \left| \hat{Y}_i - Y_i \right|
\end{equation}
where \( \hat{Y}(f) = \text{DFT}(\hat{y}(t)) \) and \( Y(f) = \text{DFT}(y(t)) \) are the predicted and true signals in the frequency domain. Similar to the time-domain loss, \textcolor{\differencecolor}{a Reward} Loss term is applied:
\begin{equation}
L_{\text{freq\_reward}} = L_{\text{freq}} + \beta \cdot \log\left(1 - e^{-L_{\text{freq}}} + \epsilon\right)
\end{equation}

\subsubsection{HTSR Loss}
The final Temporal-Spectral Reward (HTSR) loss function is a weighted combination of the time-domain and frequency-domain losses:
\begin{equation}
L_{\text{total}} = \alpha \cdot L_{\text{freq\_reward}} + (1 - \alpha) \cdot L_{\text{time\_reward}}
\end{equation}
where \( \alpha \) controls the relative importance of the frequency-domain and time-domain losses. In this work, $\alpha$ is set to $0.5$, and $\beta$ is set to $0.1$.

\subsection{Dataset}
\subsubsection{Gait-EEG Dataset}

\begin{figure}
\centering
\includegraphics[width=0.8\linewidth]{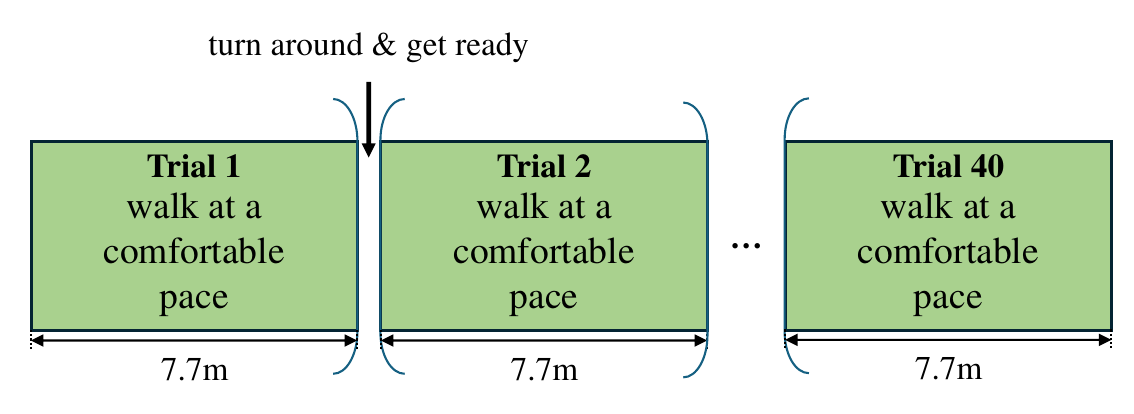}
\caption{Experiment protocol for one walking block.}
\label{fig:protocol}
\end{figure}

To investigate the \textcolor{\differencecolor}{neural mechanisms underlying natural gait}, we collected a \textcolor{\differencecolor}{novel multimodal dataset}, the Gait-EEG Dataset (GED), \textcolor{\differencecolor}{comprising synchronized recordings of brain activity and lower-limb joint kinematics during level-ground walking}. The dataset \textcolor{\differencecolor}{includes} recordings from 50 able-bodied participants (25 males, 25 females; aged 21 to 46, mean age 28.4, standard deviation 5.2), \textcolor{\differencecolor}{none of whom reported a history of neurological disorders or lower-limb pathologies}. Participants engaged in two independent level-ground walking sessions, with every session comprising three identical walking blocks. Each block included approximately 40 trials, with each trial representing EEG signals and synchronized lower-limb joint angles as the participant walked straight for 7.7 meters. The experiment protocol for each block is shown in Fig.~\ref{fig:protocol}. Sessions were spaced at least three days apart.

\begin{color}{\differencecolor}
    
\paragraph{Neural and Kinematic Data Acquisition} EEG signals were recorded using a 60-channel active EEG system (ActiCHamp Plus, Brain Products GmbH, Germany), along with 4 electrooculogram (EOG) channels to monitor ocular artifacts. Electrodes were positioned according to the extended 10–10 international system and referenced at FCz. The sampling rate was 1000 Hz. Participants wore an EEG cap and were assisted by a lightweight walking aid to reduce upper-body movement without interfering with natural gait. To further reduce motion-induced artifacts, electrode cables were secured, and participants wore tight-fitting caps and long compression pants provided by the experimenters.

Lower-limb joint angles at the hips, knees, and ankles were measured bilaterally using six twin-axis wireless goniometers (Biometrics Ltd, UK), also sampled at 1000 Hz. Sensor placement was verified through anatomical palpation at the start of each session and securely affixed using medical and kinesiology tape. The tight-fitting pants ensured both comfort during walking and minimal sensor displacement throughout the experiment.

\paragraph{Ethical Approval} This study was approved by the Institutional Review Board of Nanyang Technological University (IRB-2021-709), in accordance with applicable legislation, ethical guidelines, and safety regulations in Singapore. All participants provided written informed consent prior to participation.

\paragraph{Gait Cycle Segmentation}
The gait cycles were segmented by detecting peaks in the left knee joint angle signal recorded by the goniometer. Each local maximum peak was considered as a boundary between consecutive gait cycles. The segment of joint angle data between two consecutive peaks thus defined one gait cycle. 

\end{color}
\subsubsection{Open-access Dataset}
To further validate our proposed method, we also conducted experiments using an open-access dataset, Mobile Brain-body imaging(MoBI) dataset~\cite{he2018mobile}. The MoBI dataset includes recordings from eight participants across three sessions each. For each session, participants walked at a constant speed on a treadmill for a total of 20 minutes. During the first 15 minutes, participants received real-time feedback of their gait via an on-screen avatar mirroring their actual movements. In the final 5 minutes, the avatar’s gait feedback was driven by decoded EEG signals.

\section{Experiment}
\subsection{Data Preprocessing}\label{sec:preprocessing}
\begin{color}{\differencecolor}

\subsubsection{Conventional Preprocessing}
Prior to analysis, the EEG data from both datasets were subjected to identical preprocessing procedures. First, a minimum-phase band-pass filter (0.1--48 Hz) was applied, followed by re-referencing to the common average. The signals were then downsampled to 100 Hz. Finally, the electrooculography (EOG) channels were removed, resulting in EEG datasets with 59 channels for both the GED and MoBI datasets.

\subsubsection{Local Laplacian Filtering}
A local Laplacian filter was applied to enhance local activity within a specified area. It achieves this by computing the differences between the values of the target channel and the average of values from its neighboring channels. For each EEG channel, neighbors within a 30 mm radius were retained. This method effectively emphasizes local features while diminishing the interference from distant sources. The filtering process is expressed as:
\begin{equation}
    V'_{k} = V_{k} - \frac{1}{N} \sum_{l=1}^{N} V_{l}
\end{equation}
where \( V'_{k} \) is the Laplacian-filtered value at channel \(k\), \(V_{k}\) is the original value, \(N\) is the number of neighbors, and \(V_{l}\) are the neighboring channel values.

\subsubsection{ICA-based Artifact Removal (Control Experiment)}
To further examine the potential influence of motion- or eye-related artifacts, we conducted a control experiment using Independent Component Analysis (ICA). All independent components obtained from ICA were subjected to the computation of five established metrics \cite{mognon2011adjust}: Spatial Average Difference, Spatial Eye Difference, Generic Discontinuities Spatial Feature, Maximum Epoch Variance, and Temporal Kurtosis. 

A threshold for each metric was determined through maximum likelihood estimation. If any of the five metrics for a given component exceeded three times the threshold, that component was discarded. On average, 8.96 components were removed per session in the GED dataset and 9.54 components per session in the MoBI dataset. After artifact ICs were removed, the EOG channels were excluded, yielding a final EEG dataset of 59 channels for both datasets.
\end{color}

\subsection{Data Segregation}\label{segregation}

\subsubsection{Gait-EEG Dataset}
\begin{color}{\differencecolor}
    
Each session comprises three blocks, with each block containing 40 trials. For data partitioning, we used all trials from the first two blocks and the first 20 trials of the third block as the training set. Trials 21 to 25 of the third block were designated as the validation set, while the remaining trials (26 to 40) of the third block were concatenated along the temporal dimension to construct the test set. As EEG2GAIT is designed as a session-specific model, we independently trained an EEG2GAIT model for each session using its respective training and validation sets, and evaluated it on the corresponding session's test set.
\end{color}

\subsubsection{MoBI Dataset}
\textcolor{\differencecolor}{For each session, the first 15 minutes of walking data without closed-loop BCI control were \textcolor{\differencecolor}{allocated for model development}. Within this portion, the initial 13.5 minutes were \textcolor{\differencecolor}{used for training the model}, while the remaining 1.5 minutes (i.e., the last 10\% of this segment) were \textcolor{\differencecolor}{reserved for validation}. The final 5 minutes of walking data with closed-loop BCI control were \textcolor{\differencecolor}{kept as the independent test set}.}

\subsection{Evaluation Metric}

We evaluated the efficacy of EEG2GAIT by comparing the predicted angles of six joints with their actual recorded angles, using their correlation ($r$ value) and coefficient of determination ($R^2$ score) \textcolor{\differencecolor}{and the mean absolute error (MAE)} as the evaluation metrics: 

\begin{equation}
r=\frac{\operatorname{cov}(y, \hat{y})}{\sigma(y) \cdot \sigma(\hat{y})} 
\end{equation}

\begin{equation}
R^2=1-\frac{\sum_{i=1}^n\left(y_i-\hat{y}_i\right)^2}{\sum_{i=1}^n\left(y_i-\bar{y}\right)^2}
\end{equation}

\textcolor{\differencecolor}{
\begin{equation}
MAE = \frac{1}{n}\sum_{i=1}^n \left| y_i - \hat{y}_i \right|
\end{equation}
}

where $y$ denotes the actual joint angle, and $\hat{y}$ represents the predicted angle. The covariance between any two variables A and B is expressed as $cov(A, B)$, while $\sigma(A)$ signifies the standard deviation of A. 
Each sequence represents the data collected from a single trial, capturing the observations over a specific period. In this case, the sequence contains $n$ data points, which corresponds to $\frac{n}{100}$ seconds of data, given that the data was sampled at a frequency of 100 Hz. $\hat{y}_i$ refers to the value at the $i^{th}$ position. Lastly, $\bar{y}$ stands for the average of the actual joint angles. The $r$ value reflects the consistency in trend between the predicted and actual gait. The $R^2$ score reflects the second-order absolute error between the predicted and actual gait, providing insight into the predictive performance of the model. \textcolor{\differencecolor}{The MAE quantifies the average magnitude of the prediction error, offering a straightforward measure of how close the predictions are to the true values.}

\subsection{Implementation and Hyperparameter Settings}
The EEG2GAIT model was implemented using the PyTorch library. Training was performed using the Adam optimizer with default hyperparameter settings, and the learning rate was fixed at 0.001. A batch size of 100 was used, and training continued for a maximum of 50 epochs ($epoch_{\text{max}}$). Early stopping was applied with a patience parameter ($p$) of 30 epochs, terminating training if the Pearson correlation coefficient ($r$) on the validation set did not improve for 30 consecutive epochs.

\begin{table*}[!ht]
\centering
\caption{Prediction performance of different models on the MoBI dataset, indicated by mean $\pm$ standard deviation, averaged over the six joints. The best mean value in each column is marked in \textbf{bold}, while the best mean value in each row is marked with a superscript star (*). \textcolor{\differencecolor}{For MAE, a lower value is better ($\downarrow$)}.}
\label{tab:mobi_performance}
\vspace{1em}
\begin{color}{\differencecolor}

\textbf{$r$ value results:} \\
\vspace{0.5em}
\begin{tabular}{lc|ccc}
\toprule
\multirow{2}{*}{\textbf{Model Name\textbackslash Loss Functions}} & \multirow{2}{*}{\textbf{$L_{HTSR}$}} & \multicolumn{3}{c}{\textbf{Ablation}} \\
\cmidrule(lr){3-5}
 & & \boldmath{$L_{time+freq}$} & \boldmath{$L_{time\_reward}$} & \textbf{$L_{MSE}$} \\
\midrule
ContraWR~\cite{yang2021self} & 0.4438$^{*}\pm$0.1831 & 0.3972$\pm$0.2001 & 0.3929$\pm$0.1980 & 0.3682$\pm$0.2116 \\
FFCL~\cite{li2022motor} & 0.4460$^{*}\pm$0.1821 & 0.4284$\pm$0.1849 & 0.4277$\pm$0.1865 & 0.4002$\pm$0.1889 \\
TSception~\cite{ding2022tsception} & 0.5678$^{*}\pm$0.1733 & 0.5661$\pm$0.1725 & 0.5645$\pm$0.1731 & 0.5640$\pm$0.1731 \\
TCN~\cite{ingolfsson2020eeg} & 0.6040$^{*}\pm${0.1541} & 0.5985$\pm${0.1538} & 0.5934$\pm${0.1532} & 0.5934$\pm${0.1541} \\
ST-Transformer~\cite{song2021transformer} & 0.7171$^{*}\pm$0.1724 & 0.7085$\pm$0.1665 & 0.7063$\pm$0.1769 & 0.6954$\pm$0.1760 \\
EEGConformer~\cite{song2023eeg} & 0.6844$^{*}\pm$0.1830 & 0.6770$\pm$0.1876 & 0.6712$\pm$0.1883 & 0.6651$\pm$0.1896 \\
SPaRCNet~\cite{jing2023development} & 0.6838$^{*}\pm$0.1973 & 0.6671$\pm$0.2005 & 0.6709$\pm$0.1998 & 0.6623$\pm$0.1981 \\
EEGNet~\cite{lawhern2018eegnet} & 0.6898$^{*}\pm$0.1788 & 0.6768$\pm$0.1751 & 0.6747$\pm$0.1858 & 0.6740$\pm$0.1806 \\
deepConvNet~\cite{schirrmeister2017deep} & 0.7588$^{*}\pm$0.1819 & 0.7449$\pm$0.1950 & 0.7453$\pm$0.1951 & 0.7356$\pm$0.1972 \\
EEG2GAIT(w/o HGP) & 0.7629$^{*}\pm$0.1751 & 0.7499$\pm$0.1840 & 0.7531$\pm$0.1814 & 0.7380$\pm$0.1901 \\
EEG2GAIT & \textbf{0.7793}$^{*}\pm$0.1630 & \textbf{0.7688}$\pm$0.1676 & \textbf{0.7691}$\pm$0.1674 & \textbf{0.7618}$\pm$0.1694 \\
\bottomrule
\end{tabular}

\vspace{1em}

\textbf{$R^2$ score results:} \\
\vspace{0.5em}
\begin{tabular}{lc|ccc}
\toprule
\multirow{2}{*}{\textbf{Model Name\textbackslash Loss Functions}} & \multirow{2}{*}{\textbf{$L_{HTSR}$}} & \multicolumn{3}{c}{\textbf{Ablation}} \\
\cmidrule(lr){3-5}
 & & \boldmath{$L_{time+freq}$} & \boldmath{$L_{time\_reward}$} & \textbf{$L_{MSE}$} \\
\midrule
ContraWR~\cite{yang2021self} & -0.0021$^{*}\pm$0.3021 & -0.0101$\pm$0.2613 & -0.0272$\pm$0.2790 & -0.0363$\pm$0.2789 \\
FFCL~\cite{li2022motor} & 0.0337$^{*}\pm$0.2629 & 0.0306$\pm$0.2782 & 0.0055$\pm$0.2710 & -0.0028$\pm$0.2597 \\
TSception~\cite{ding2022tsception} & 0.2342$^{*}\pm$0.2877 & 0.2298$\pm$0.2869 & 0.2259$\pm$0.2920 & 0.2288$\pm$0.2916 \\
TCN~\cite{ingolfsson2020eeg} & 0.3492$^{*}\pm${0.2057} & 0.3396$\pm${0.2054} & 0.3359$\pm${0.1976} & 0.3371$\pm${0.1986} \\
ST-Transformer~\cite{song2021transformer} & 0.4851$^{*}\pm$0.2649 & 0.4702$\pm$0.2579 & 0.4721$\pm$0.2624 & 0.4530$\pm$0.2626 \\
EEGConformer~\cite{song2023eeg} & 0.4571$^{*}\pm$0.2521 & 0.4444$\pm$0.2603 & 0.4375$\pm$0.2590 & 0.4268$\pm$0.2605 \\
SPaRCNet~\cite{jing2023development} & 0.4188$^{*}\pm$0.3202 & 0.3938$\pm$0.3225 & 0.3978$\pm$0.3279 & 0.3851$\pm$0.3208 \\
EEGNet~\cite{lawhern2018eegnet} & 0.4574$^{*}\pm$0.2312 & 0.4336$\pm$0.2248 & 0.4315$\pm$0.2420 & 0.4338$\pm$0.2341 \\
deepConvNet~\cite{schirrmeister2017deep} & 0.5585$^{*}\pm$0.2941 & 0.5355$\pm$0.3096 & 0.5365$\pm$0.3096 & 0.5236$\pm$0.2995 \\
EEG2GAIT(w/o HGP) & 0.5720$^{*}\pm$0.2673 & 0.5502$\pm$0.2816 & 0.5563$\pm$0.2801 & 0.5278$\pm$0.2970 \\
EEG2GAIT & \textbf{0.5965}$^{*}\pm$0.2490 & \textbf{0.5764}$\pm$0.2606 & \textbf{0.5803}$\pm$0.2559 & \textbf{0.5662}$\pm$0.2598 \\
\bottomrule
\end{tabular}

\vspace{1em}

\textbf{MAE results ($\downarrow$):} \\
\vspace{0.5em}
\begin{tabular}{lc|ccc}
\toprule
\multirow{2}{*}{\textbf{Model Name\textbackslash Loss Functions}} & \multirow{2}{*}{\textbf{$L_{HTSR}$}} & \multicolumn{3}{c}{\textbf{Ablation}} \\
\cmidrule(lr){3-5}
 & & \boldmath{$L_{time+freq}$} & \boldmath{$L_{time\_reward}$} & \textbf{$L_{MSE}$} \\
\midrule
ContraWR~\cite{yang2021self} & 7.7427$^{*}\pm$1.6476 & 7.8940$\pm$1.5306 & 7.9890$\pm$1.6028 & 8.0734$\pm$1.5231 \\
FFCL~\cite{li2022motor} & 7.6539$^{*}\pm$1.5389 & 7.7645$\pm$1.5822 & 7.8065$\pm$1.5637 & 7.9373$\pm$1.4887 \\
TSception~\cite{ding2022tsception} & 7.1113$^{*}\pm$1.4242 & 7.1251$\pm$1.4164 & 7.1332$\pm$1.4341 & 7.1457$\pm$1.4206 \\
TCN~\cite{ingolfsson2020eeg} & 6.1885$^{*}\pm${1.1872} & 6.2294$\pm${1.1646} & 6.2554$\pm${1.1562} & 6.2486$\pm${1.1676} \\
ST-Transformer~\cite{song2021transformer} & 4.9430$^{*}\pm$1.6034 & 5.0568$\pm$1.5896 & 5.0447$\pm$1.6287 & 5.1566$\pm$1.6480 \\
EEGConformer~\cite{song2023eeg} & 5.3537$^{*}\pm$1.5794 & 5.4356$\pm$1.6406 & 5.4647$\pm$1.6399 & 5.5074$\pm$1.6185 \\
SPaRCNet~\cite{jing2023development} & 5.3655$^{*}\pm$1.8377 & 5.5026$\pm$1.8355 & 5.4351$\pm$1.8549 & 5.5389$\pm$1.8055 \\
EEGNet~\cite{lawhern2018eegnet} & 5.7088$^{*}\pm$1.3842 & 5.8318$\pm$1.3622 & 5.8668$\pm$1.4331 & 5.8418$\pm$1.3992 \\
deepConvNet~\cite{schirrmeister2017deep} & 4.6303$^{*}\pm$1.8319 & 4.7507$\pm$1.9069 & 4.7433$\pm$1.9068 & 4.8539$\pm$1.8867 \\
EEG2GAIT(w/o HGP) & 4.5107$^{*}\pm$1.7550 & 4.6413$\pm$1.8100 & 4.5986$\pm$1.8054 & 4.7667$\pm$1.8786 \\
EEG2GAIT & \textbf{4.3836}$^{*}\pm$1.6655 & \textbf{4.4839}$\pm$1.6963 & \textbf{4.4574}$\pm$1.6872 & \textbf{4.5494}$\pm$1.7088 \\
\bottomrule
\end{tabular}
    
\end{color}
\end{table*}
\section{Results and Analysis}
\begin{figure}
\centering
\includegraphics[width=\linewidth]{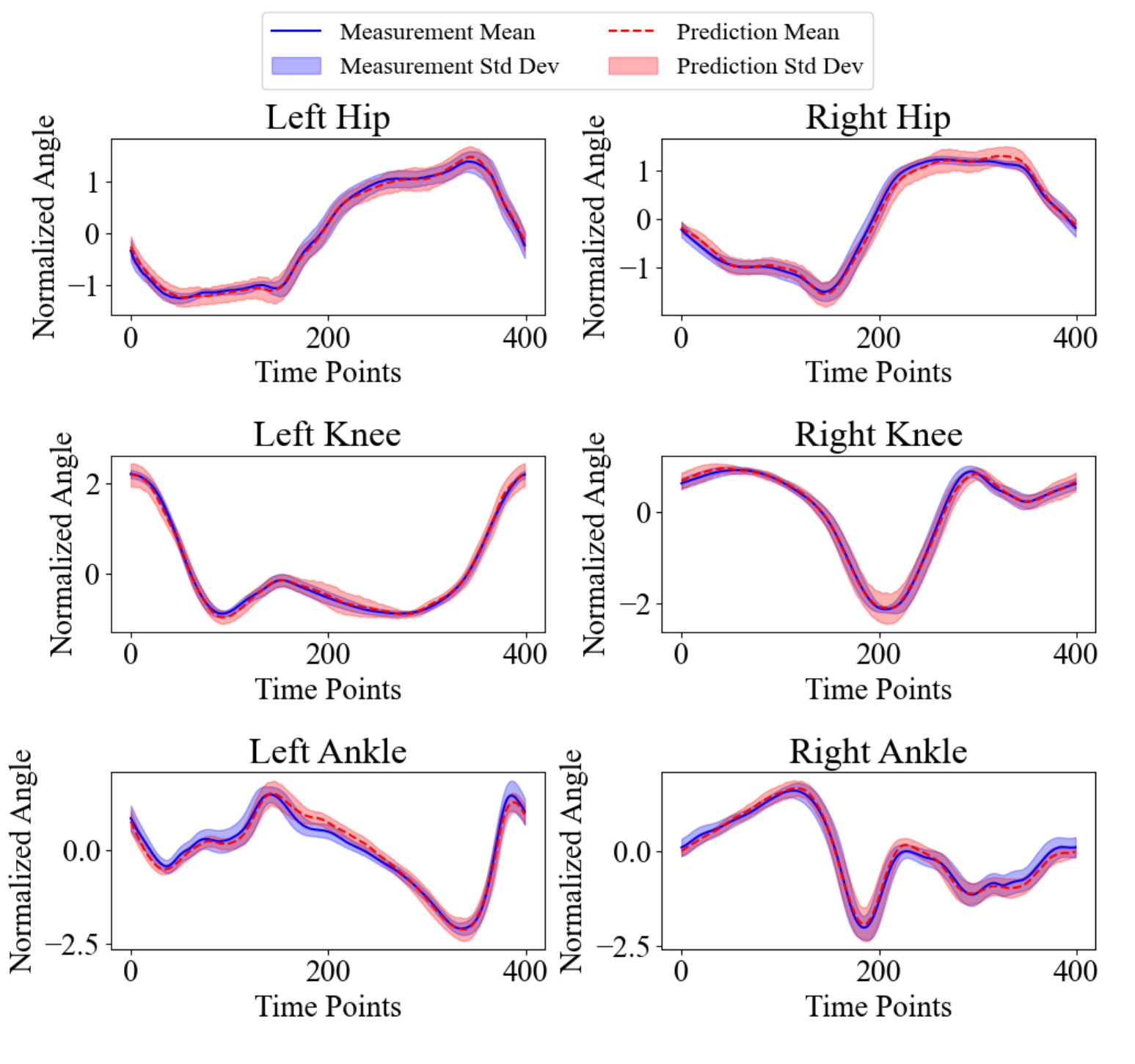}
\caption{\textcolor{\differencecolor}{Visualization of the actual and predicted six joint angles for the test set in participant 2's second session. The goniometer-measured angles were segmented into gait cycles using peak detection, with each gait cycle interpolated to 400 time points. The blue solid line represents the actual mean angle, and the blue shaded area indicates the standard deviation of the measured angles across all gait cycles. The red dashed line denotes the predicted mean angle, while the red shaded area shows the standard deviation of the predictions. }
}\label{fig:periodic_vis}
\end{figure}
In this section, we validate the performance of the proposed method on the Gait-EEG dataset and MoBI dataset~\cite{he2018mobile} and compared with several state-of-the-art deep learning and machine learning algorithms in the brain-computer interface (BCI) domain, including ContraWR~\cite{yang2021self}, FFCL~\cite{li2022motor}, TSception~\cite{ding2022tsception}, Temporal Convolutional Network (TCN)~\cite{ingolfsson2020eeg}, ST-Transformer~\cite{song2021transformer}, EEGConformer~\cite{song2023eeg}, SPaRCNet~\cite{jing2023development}, EEGNet~\cite{lawhern2018eegnet}, and deepConvNet~\cite{schirrmeister2017deep}. 
An ablation study was conducted to demonstrate the contributions of the HGP module and HTSR loss. Saliency mapping across EEG channels identified specific channels with notably higher contributions to the decoding task, offering a spatial interpretation of the neural correlates involved in lower-limb motor decoding as well as validating the obtained features. This confirms that the model effectively leverages signals from motor-related areas. Additionally, we illustrate the overlay of decoding results across different models for the gait cycle, offering a more intuitive display of EEG2GAIT's superior performance.

Fig. \textcolor{black}{\ref{fig:periodic_vis}} illustrates the visualization of the predicted and actual six joint angles for the left knee during the test set of participant 2's second session \textcolor{black}{($r$ = \textcolor{\differencecolor}{$0.978$}, $R^2$ = \textcolor{\differencecolor}{$0.955$})}. To standardize the input for analysis and modeling, each gait cycle was temporally normalized by interpolating the angle data to a fixed length of 400 time points, enabling consistent comparison across different cycles and participants.

The blue solid line represents the mean goniometer-measured angle for each time point across all gait cycles, while the blue shaded region indicates the standard deviation of the measured values, reflecting variability across gait cycles. The red dashed line shows the predicted mean knee joint angle at each time point, and the red shaded region depicts the predicted standard deviation. 

\begin{table*}[ht]
\centering
\caption{
Prediction performance of different models on GED, indicated by mean $\pm$ standard deviation, averaged over the six joints. The best mean value in each column is marked in \textbf{bold}, while the best mean value in each row is marked with a superscript star (*). \textcolor{\differencecolor}{For MAE, a lower value is better ($\downarrow$)}.
}
\label{tab:performance_comparison}
\vspace{1em}
\begin{color}{\differencecolor}
\textbf{$r$ value results:} \\ 
\vspace{0.5em}
\begin{tabular}{lc|ccc}
\toprule
\multirow{2}{*}{\textbf{Model Name\textbackslash Loss Functions}} & \multirow{2}{*}{\textbf{$L_{HTSR}$}} & \multicolumn{3}{c}{\textbf{Ablation}} \\
\cmidrule(lr){3-5}
 & & \boldmath{$L_{time+freq}$} & \boldmath{$L_{time\_reward}$} & \textbf{$L_{MSE}$} \\
\midrule
ContraWR~\cite{yang2021self} & 0.6076$^{*}\pm$0.2589 & 0.5737$\pm$0.2792 & 0.5870$\pm$0.2654 & 0.5427$\pm$0.2849 \\
FFCL~\cite{li2022motor} & 0.6473$^{*}\pm$0.1982 & 0.6299$\pm$0.2048 & 0.6293$\pm$0.2051 & 0.5959$\pm$0.2137 \\
TSception~\cite{ding2022tsception} & 0.7214$^{*}\pm$0.1175 & 0.7116$\pm$0.1184 & 0.7059$\pm$0.1158 & 0.6943$\pm$0.1198 \\
TCN~\cite{ingolfsson2020eeg} & 0.7335$^{*}\pm$0.1439 & 0.7249$\pm$0.1464 & 0.7202$\pm$0.1483 & 0.7092$\pm$0.1532 \\
ST-Transformer~\cite{song2021transformer} & 0.9193$^{*}\pm$0.0613 & 0.9162$\pm$0.0632 & {0.9155}$\pm$0.0623 & 0.9110$\pm$0.0649 \\
EEGConformer~\cite{song2023eeg} & 0.9001$^{*}\pm$0.0706 & 0.8966$\pm$0.0727 & 0.8967$\pm$0.0717 & 0.8918$\pm$0.0746 \\
SPaRCNet~\cite{jing2023development} & 0.8852$^{*}\pm$0.0899 & 0.8777$\pm$0.0960 & 0.8751$\pm$0.0965 & 0.8712$\pm$0.0994 \\
EEGNet~\cite{lawhern2018eegnet} & 0.8885$^{*}\pm$0.0681 & 0.8847$\pm$0.0707 & 0.8806$\pm$0.0743 & 0.8715$\pm$0.0803 \\
deepConvNet~\cite{schirrmeister2017deep} & 0.9497$^{*}\pm$0.0365 & 0.9447$\pm$0.0395 & 0.9444$\pm$0.0384 & 0.9414$\pm$0.0388 \\
EEG2GAIT(w/o HGP) & 0.9526$^{*}\pm$0.0350 & 0.9483$\pm$0.0367 & 0.9499$\pm$0.0340 & 0.9452$\pm$0.0376 \\
EEG2GAIT & \textbf{0.9588}$^{*}\pm$0.0307 & \textbf{0.9558}$\pm$0.0358 & \textbf{0.9551}$\pm$0.0390 & \textbf{0.9467}$\pm$0.0912 \\
\bottomrule
\end{tabular}

\vspace{1em}

\textbf{$R^2$ score results:} \\
\vspace{0.5em}
\begin{tabular}{lc|ccc}
\toprule
\multirow{2}{*}{\textbf{Model Name\textbackslash Loss Functions}} & \multirow{2}{*}{\textbf{$L_{HTSR}$}} & \multicolumn{3}{c}{\textbf{Ablation}} \\
\cmidrule(lr){3-5}
 & & \boldmath{$L_{time+freq}$} & \boldmath{$L_{time\_reward}$} & \textbf{$L_{MSE}$} \\
\midrule
ContraWR~\cite{yang2021self} & 0.3058$^{*}\pm$0.3264 & 0.2906$\pm$0.3097 & 0.2866$\pm$0.3239 & 0.2676$\pm$0.3014 \\
FFCL~\cite{li2022motor} & 0.3666$^{*}\pm$0.3192 & 0.3522$\pm$0.3179 & 0.3503$\pm$0.3190 & 0.3104$\pm$0.3098 \\
TSception~\cite{ding2022tsception} & 0.4560$^{*}\pm$0.3426 & 0.4403$\pm$0.3377 & 0.4293$\pm$0.3345 & 0.4060$\pm$0.3717 \\
TCN~\cite{ingolfsson2020eeg} & 0.5244$^{*}\pm$0.2613 & 0.5111$\pm$0.2712 & 0.5067$\pm$0.2647 & 0.4902$\pm$0.2731 \\
ST-Transformer~\cite{song2021transformer} & 0.8219$^{*}\pm$0.2206 & 0.8188$\pm$0.2008 & 0.8154$\pm$0.2111 & 0.8095$\pm$0.1941 \\
EEGConformer~\cite{song2023eeg} & 0.7901$^{*}\pm$0.2184 & 0.7809$\pm$0.2419 & 0.7813$\pm$0.2384 & 0.7773$\pm$0.2075 \\
SPaRCNet~\cite{jing2023development} & 0.7532$^{*}\pm$0.2934 & 0.7396$\pm$0.2948 & 0.7405$\pm$0.2572 & 0.7334$\pm$0.2591 \\
EEGNet~\cite{lawhern2018eegnet} & 0.7604$^{*}\pm$0.1846 & 0.7497$\pm$0.2152 & 0.7470$\pm$0.1922 & 0.7303$\pm$0.2023 \\
deepConvNet~\cite{schirrmeister2017deep} & 0.8769$^{*}\pm$0.2251 & 0.8650$\pm$0.2349 & 0.8645$\pm$0.2328 & 0.8527$\pm$0.2717 \\
EEG2GAIT(w/o HGP) & 0.9008$^{*}\pm$0.0723 & 0.8774$\pm$0.1887 & 0.8787$\pm$0.1959 & 0.8672$\pm$0.2221 \\
EEG2GAIT & \textbf{0.9141}$^{*}\pm$0.0627 & \textbf{0.9085}$\pm$0.0693 & \textbf{0.9068}$\pm$0.0771 & \textbf{0.8903}$\pm$0.1782 \\
\bottomrule
\end{tabular}

\vspace{1em}

\textbf{MAE results ($\downarrow$):} \\
\vspace{0.5em}
\begin{tabular}{lc|ccc}
\toprule
\multirow{2}{*}{\textbf{Model Name\textbackslash Loss Functions}} & \multirow{2}{*}{\textbf{$L_{HTSR}$}} & \multicolumn{3}{c}{\textbf{Ablation}} \\
\cmidrule(lr){3-5}
 & & \boldmath{$L_{time+freq}$} & \boldmath{$L_{time\_reward}$} & \textbf{$L_{MSE}$} \\
\midrule
ContraWR~\cite{yang2021self} & 0.5875$^{*}\pm$0.1619 & 0.6032$\pm$0.1621 & 0.6029$\pm$0.1634 & 0.6250$\pm$0.1622 \\
FFCL~\cite{li2022motor} & 0.5579$^{*}\pm$0.1569 & 0.5731$\pm$0.1535 & 0.5744$\pm$0.1548 & 0.5979$\pm$0.1544 \\
TSception~\cite{ding2022tsception} & 0.5592$^{*}\pm$0.1064 & 0.5683$\pm$0.1070 & 0.5759$\pm$0.1047 & 0.5859$\pm$0.1055 \\
TCN~\cite{ingolfsson2020eeg} & 0.4887$^{*}\pm$0.1326 & 0.4958$\pm$0.1325 & 0.5029$\pm$0.1324 & 0.5096$\pm$0.1345 \\
ST-Transformer~\cite{song2021transformer} & 0.2630$^{*}\pm$0.0840 & 0.2645$\pm$0.0855 & 0.2665$\pm$0.0837 & 0.2702$\pm$0.0861 \\
EEGConformer~\cite{song2023eeg} & 0.2925$^{*}\pm$0.0888 & 0.2965$\pm$0.0898 & 0.2978$\pm$0.0898 & 0.3031$\pm$0.0913 \\
SPaRCNet~\cite{jing2023development} & 0.3077$^{*}\pm$0.1076 & 0.3170$\pm$0.1103 & 0.3195$\pm$0.1111 & 0.3237$\pm$0.1125 \\
EEGNet~\cite{lawhern2018eegnet} & 0.3515$^{*}\pm$0.0917 & 0.3573$\pm$0.0955 & 0.3634$\pm$0.0958 & 0.3746$\pm$0.0989 \\
deepConvNet~\cite{schirrmeister2017deep} & 0.2125$^{*}\pm$0.0709 & 0.2287$\pm$0.0774 & 0.2288$\pm$0.0769 & 0.2373$\pm$0.0777 \\
EEG2GAIT(w/o HGP) & 0.2056$^{*}\pm$0.0651 & 0.2160$\pm$0.0694 & 0.2136$\pm$0.0669 & 0.2230$\pm$0.0693 \\
EEG2GAIT & \textbf{0.1930}$^{*}\pm$0.0622 & \textbf{0.1979}$\pm$0.0652 & \textbf{0.1988}$\pm$0.0662 & \textbf{0.2059}$\pm$0.0830 \\
\bottomrule
\end{tabular}
\end{color}
\end{table*}

\subsection{Model Performance}
The results of all metrics on the test sets are presented to evaluate the performance of EEG2GAIT in comparison to baseline methods on the MoBI and GED dataset. EEG2GAIT achieved an $r$ value of \textcolor{\differencecolor}{0.959} ($R^2$ = \textcolor{\differencecolor}{0.914}) on the GED dataset with 50 subjects and an $r$ value of \textcolor{\differencecolor}{0.779} ($R^2$ = \textcolor{\differencecolor}{0.597}) on the MoBI dataset with 8 subjects, outperforming the baseline methods. All baseline methods were trained according to the strategy described in Section \ref{segregation}. A summary of the results for EEG2GAIT and the baseline methods on the different datasets is provided in Table \textcolor{\differencecolor}{\ref{tab:mobi_performance}} and \textcolor{black}{\ref{tab:performance_comparison}}. Overall, EEG2GAIT demonstrated the best performance across all metrics among all methods, with a lower standard deviation than most methods. \textcolor{\differencecolor}{In addition, the decoding performance of EEG2GAIT for each individual joint across the three evaluation metrics ($r$, $R^2$, and MAE) is reported in Table \ref{tab:joint_specific_results}.}

\begin{table}[htbp]
\begin{color}{\differencecolor}
  
    \centering
    \caption{\textcolor{\differencecolor}{Joint-specific prediction results of EEG2GAIT model on GED across six lower-limb joints, indicated by mean $\pm$ standard deviation}}
    \label{tab:joint_specific_results}
    \vspace{0.2cm}
   \begin{tabular}{lccc}
\hline
\textbf{Joint} & \textbf{$r$ value} & \textbf{$R^2$ score} & \textbf{MAE} \\
\hline
Left Hip   & 0.9677 $\pm$ 0.0302 & 0.9309 $\pm$ 0.0614 & 0.1761 $\pm$ 0.0741 \\
Left Knee  & 0.9729 $\pm$ 0.0256 & 0.9437 $\pm$ 0.0522 & 0.1556 $\pm$ 0.0660 \\
Left Ankle & 0.9367 $\pm$ 0.0453 & 0.8669 $\pm$ 0.0902 & 0.2528 $\pm$ 0.0796 \\
Right Hip  & 0.9674 $\pm$ 0.0348 & 0.9313 $\pm$ 0.0686 & 0.1744 $\pm$ 0.0639 \\
Right Knee & 0.9726 $\pm$ 0.0262 & 0.9391 $\pm$ 0.0768 & 0.1490 $\pm$ 0.0613 \\
Right Ankle& 0.9353 $\pm$ 0.0441 & 0.8730 $\pm$ 0.0801 & 0.2502 $\pm$ 0.0718 \\
\hline
\end{tabular}

  \end{color}  
    
    \end{table}

\subsection{Ablation Analysis}

The results of the loss ablation studies can be observed across the columns in Table \textcolor{\differencecolor}{\ref{tab:mobi_performance} and} \ref{tab:performance_comparison}. \textcolor{\differencecolor}{To assess statistical significance, we tested whether the performance improvements of HTSR loss over MSE loss were greater than zero. Results confirmed that} \textcolor{\differencecolor}{HTSR loss produced significantly higher performance} across all baseline models and our EEG2GAIT model ($p \ll 0.001$ \textcolor{\differencecolor}{for all three metrics, one-tailed paired-sample t-test}). Furthermore, as we progressively isolated each component of the loss function, the results showed that each individual component contributed positively to model performance for most models.

\textcolor{\differencecolor}{To evaluate the contribution of the HGP module, we conducted an ablation study comparing EEG2GAIT with and without HGP. On the GPP dataset, the inclusion of HGP significantly improved performance across all metrics ($p \ll 0.001$). On the MoBI dataset, the improvements were also statistically significant, with $p=3.34\times10^{-3}$ for $r$, and $p=1.13\times10^{-2}$ for $R^{2}$ and $p=3.42\times10^{-2}$ for MAE. Even without HGP, EEG2GAIT still outperformed all baseline models, highlighting the robustness of the core framework, while the inclusion of HGP provided consistent and statistically significant gains.}

\begin{color}{\differencecolor}

\subsection{Impact of ICA-based Artifact Removal}
We trained all models on the ICA-cleaned datasets following the same procedures described in Section~\ref{sec:preprocessing}. 
The results are summarized in Table~\ref{tab:ica_ablation}. 
EEG2GAIT achieved a high decoding accuracy after ICA cleaning, with an average $r$ value of 0.935, slightly lower than its performance on the unprocessed data. 

For baseline methods, the effect of ICA cleaning was mixed: several models (e.g., ContraWR, FFCL, TSception, SPaRCNet) showed performance gains after ICA, whereas others (e.g., ST-Transformer, EEGConformer, EEGNet, deepConvNet) exhibited decreases.

\begin{table*}[!ht]
\begin{color}{\differencecolor}
\centering
\caption{
\textcolor{\differencecolor}{Performance comparison on GED and MoBI datasets with and without ICA-based artifact removal. 
Values are reported as mean across participants. 
For each metric, the single best value across all models and ICA settings is highlighted in \textbf{bold}. 
Within each model, the better of w/o ICA vs w/ ICA is marked with $^{*}$. 
Values are presented in the order: \textbf{without ICA / with ICA}.}
}
\label{tab:ica_ablation}
\resizebox{\textwidth}{!}{
\begin{tabular}{lcccccc}
\toprule
\multirow{2}{*}{\textbf{Model}} & 
\multicolumn{3}{c}{\textbf{GED}} & 
\multicolumn{3}{c}{\textbf{MoBI}} \\
\cmidrule(lr){2-4} \cmidrule(lr){5-7}
& $r$ value & $R^2$ score & MAE ($\downarrow$) & $r$ value & $R^2$ score & MAE ($\downarrow$) \\
\midrule
ContraWR~\cite{yang2021self} & 0.6076 / 0.6263$^{*}$ & 0.3058 / 0.3142$^{*}$ & 0.5875 / 0.5698$^{*}$ & 0.4438$^{*}$ / 0.3140 & -0.0021 / 0.0427$^{*}$ & 7.74 / 7.60$^{*}$ \\
FFCL~\cite{li2022motor} & 0.6473 / 0.6710$^{*}$ & 0.3666 / 0.4046$^{*}$ & 0.5579 / 0.5563$^{*}$ & 0.4460$^{*}$ / 0.4407 & 0.0337 / 0.1413$^{*}$ & 7.65 / 7.23$^{*}$ \\
TSception~\cite{ding2022tsception} & 0.7214 / 0.7257$^{*}$ & 0.4560 / 0.4813$^{*}$ & 0.5592 / 0.5451$^{*}$ & 0.5678$^{*}$ / 0.5527 & 0.2342$^{*}$ / 0.1918 & 7.11$^{*}$ / 7.33 \\
TCN~\cite{ingolfsson2020eeg} & 0.7335$^{*}$ / 0.7034 & 0.5244$^{*}$ / 0.4914 & 0.4887$^{*}$ / 0.5351 & 0.6040 / 0.6099$^{*}$ & 0.3492$^{*}$ / 0.3423 & 6.19$^{*}$ / 6.37 \\
ST-Transformer~\cite{song2021transformer} & 0.9193$^{*}$ / 0.8898 & 0.8219$^{*}$ / 0.7681 & 0.2630$^{*}$ / 0.3153 & 0.7171$^{*}$ / 0.6760 & 0.4851$^{*}$ / 0.4099 & 4.94$^{*}$ / 5.92 \\
EEGConformer~\cite{song2023eeg} & 0.9001$^{*}$ / 0.8739 & 0.7901$^{*}$ / 0.7597 & 0.2925$^{*}$ / 0.3226 & 0.6844$^{*}$ / 0.6027 & 0.4571$^{*}$ / 0.3325 & 5.35$^{*}$ / 6.27 \\
SPaRCNet~\cite{jing2023development} & 0.8852 / 0.8877$^{*}$ & 0.7532 / 0.7534$^{*}$ & 0.3077$^{*}$ / 0.3120 & 0.6838$^{*}$ / 0.6459 & 0.4188$^{*}$ / 0.3586 & 5.37$^{*}$ / 6.38 \\
EEGNet~\cite{lawhern2018eegnet} & 0.8885$^{*}$ / 0.8641 & 0.7604$^{*}$ / 0.7150 & 0.3515$^{*}$ / 0.3642 & 0.6898$^{*}$ / 0.6495 & 0.4574$^{*}$ / 0.3988 & 5.71$^{*}$ / 5.74 \\
deepConvNet~\cite{schirrmeister2017deep} & 0.9497$^{*}$ / 0.9224 & 0.8769$^{*}$ / 0.8375 & 0.2125$^{*}$ / 0.2504 & 0.7588$^{*}$ / 0.6854 & 0.5585$^{*}$ / 0.4539 & 4.63$^{*}$ / 5.49 \\
EEG2GAIT (w/o HGP) & 0.9526$^{*}$ / 0.9257 & 0.9008$^{*}$ / 0.8487 & 0.2056$^{*}$ / 0.2183 & 0.7629$^{*}$ / 0.7009 & 0.5720$^{*}$ / 0.4517 & 4.51$^{*}$ / 5.61 \\
EEG2GAIT & \textbf{0.9588}$^{*}$ / 0.9351 & \textbf{0.9141}$^{*}$ / 0.8651 & \textbf{0.1930}$^{*}$ / 0.2107 & \textbf{0.7793}$^{*}$ / 0.7126 & \textbf{0.5965}$^{*}$ / 0.4835 & \textbf{4.38}$^{*}$ / 4.85 \\
\bottomrule
\end{tabular}}
\end{color}
\end{table*}

\end{color}

\subsection{Spatial Analysis}
In addition to the performance metrics presented, we expanded our evaluation to include spatial feature importance analysis to better understand critical areas during decoding. To achieve this, we employed saliency mapping—a technique in machine learning that visualizes the importance of each input feature for the model's predictions~\cite{simonyan2013deep}. This method highlights the input areas the model is most sensitive to when making predictions. The saliency map, $S$, is generated by calculating the gradient of the model’s output with respect to each input feature. The gradients are visualized to represent how variations in each input element, $X_{ij}$ (where $i, j$ are the spatial and temporal indices of $X$ , an input sample in $\mathbf{R}^{C\times T}$), influence the output prediction. The magnitude of each element $S_{ij}$ in $S$ illustrates the importance of the corresponding input pixel $X_{ij}$ to the output prediction.

To derive a spatial saliency map from these calculations, we first averaged $S$ across the temporal dimension to obtain \( \Bar{S} \). We then projected \( \Bar{S} \) onto the corresponding scalp electrode positions by averaging across all time points of $S$, creating a topographical map that illustrates the focal areas of brain activity relevant to the model's decisions.

\begin{figure*}[htb]
  
    \centering
    \includegraphics[width=0.9\linewidth]{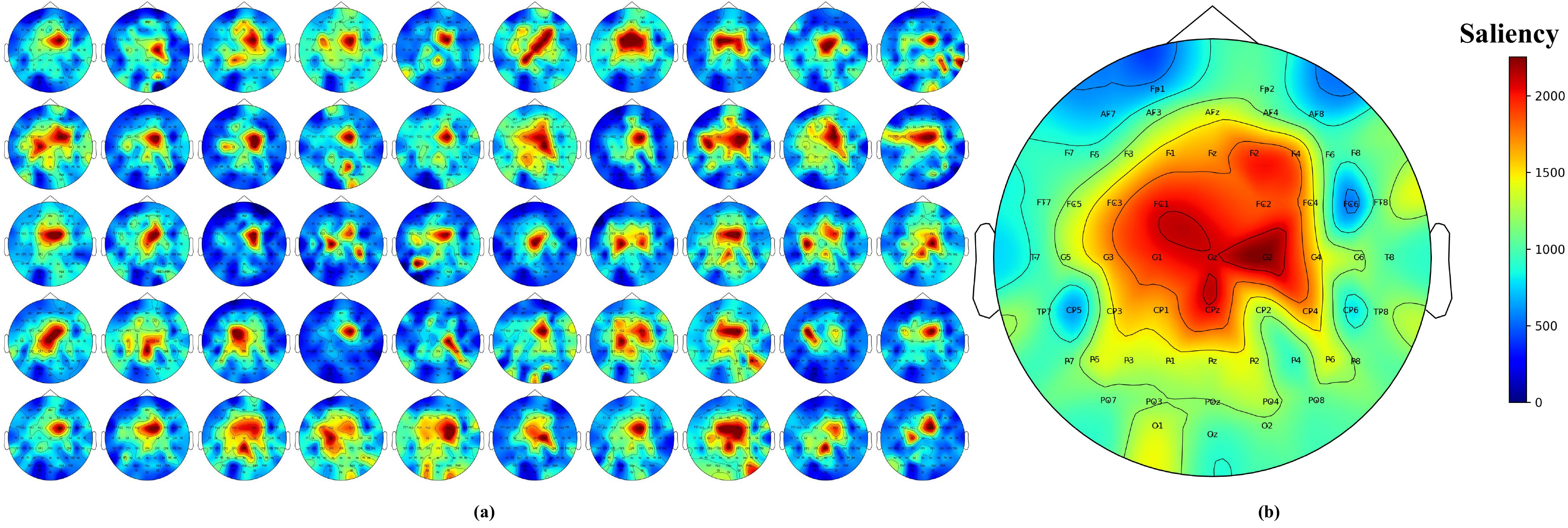}
    \caption{Saliency maps generated by EEG2GAIT: \textbf{(a)} individual saliency maps for each subject, and \textbf{(b)} the average saliency map computed across all subjects.}

  \label{fig:saliency_inner}
\end{figure*}

The saliency maps are plotted in Figure \ref{fig:saliency_inner}. The results indicated that the highlighted EEG channels during subjects' walking were concentrated in the motor areas, especially in the lower-limb area, which is the anterior paracentral area~\cite{penfield1937somatic}. Our observations consistently pointed towards the motor areas, particularly channel Cz, FC1, FC2, Fz, CP1, CP2, \textcolor{\differencecolor}{CPz}, exhibiting notably higher saliency compared to other regions. As shown in Figure \ref{fig:saliency_inner}, the average saliency maps across all subjects indicate strong role of central motor region in gait pattern encoding.

\section{Discussion}
In our current work, we proposed EEG2GAIT, a novel hierarchical graph neural network model specifically designed to address the limitations of existing approaches, which often fail to fully explore neurophysiological priors, thereby neglecting the spatial relationships between individual recording channels. Comparisons with state-of-the-art models demonstrated that EEG2GAIT achieves superior performance on two independent gait datasets. A novel loss function was introduced to leverage the temporal and spectral information inherent in EEG recordings and mitigate the limitations of existing loss functions. Feature importance analyses finally highlighted the considerable contributions of cortical regions known to be involved in gait generation, thereby providing neurophysiological validation for the obtained results.

\subsection{Hierarchical Graph Convolutional Network Pyramid}

The proposed Hierarchical Graph Convolutional Network Pyramid (HGP) introduced a novel framework for capturing spatial relationships among EEG channels associated with gait-related brain activity by modeling EEG signals as a non-Euclidean structure. Direct comparison with existing state-of-the-art models in EEG-decoding revealed superior performance of the proposed approach, both in terms of achieved correlation and $R^2$-score between prediction and ground truth, and in the corresponding standard deviations (see Table \ref{tab:performance_comparison}). Importantly, the use of hierarchical graph convolutional layers with learnable adjacency matrices displayed the best regression performance of all compared methods, showing slight improvement compared to the basic EEG2Gait model without the hierarchical graph convolutional network (GCN) pyramid (HGP), specifically in the $R^2$-score. These findings stands in line with previous research, which documented effective motor activity decoding during gait by leveraging hierarchical graph convolutional layers with learnable adjacency matrices~\cite{sun2021adaptive, fu2021decoding}.


The reported performance of EEG2GAIT in aligning predicted and actual gait sequences within different datasets (see Tables \ref{tab:mobi_performance} and \ref{tab:performance_comparison}) highlights its ability to model the neural underpinnings of gait with high fidelity. Specifically, the proposed model largely outperformed nine other state-of-the-art benchmark models specifically designed to decode information from ongoing EEG signals in terms of mean squared error between regressed and actual gait signals. As demonstrated in the ablation study (Table \ref{tab:performance_comparison}), we also compared the performance of EEG2GAIT without HGP. The results indicate that with the inclusion of HGP, EEG2GAIT achieves slightly enhanced performance in both the $r$ value and $R^2$ score dimensions, demonstrating that the addition of HGP consistently improves EEG2GAIT's performance across all loss functions. Furthermore, the reduction in variance indicates that the inclusion of HGP makes the model's performance more stable.

\subsection{Combined temporal and spectral losses}
The proposed Hybrid Temporal-Spectral Reward (HTSR) Loss addresses key challenges in continuous gait decoding from EEG by combining time-domain, frequency-domain, and reward mechanisms. As shown in Table \ref{tab:performance_comparison}, the use of time- and frequency-domain information (second column) consistently improved both correlations and $R^2$-score across the regarded models when only using a MSE loss (first column). Likewise, consistent performance improvements emerged when rewarding accurate inferences via a temporal reward loss (third column) compared to an ordinary MSE loss (first column), indicating that learning from correct examples, as opposed to the traditional approach of learning via penalties, indeed improves the models performance. Interestingly, combining the two approaches to our proposed hybrid temporal-spectral reward (HTSR) loss yielded superior performance across all models. The performance improvement most likely stems from the fact that the hybrid framework enhances the model’s ability to capture subtle temporal patterns and long-term contextual information for decoding. Specifically, the time-domain loss ensures accurate gait prediction, while the reward mechanism preserves fine-grained temporal details. Additionally, the frequency-domain component, leveraging the Discrete Fourier Transform (DFT), effectively captures long-term dependencies, with a reward term that emphasizes key spectral features. Overall, the HTSR loss outperforms the conventional MSE loss function, demonstrating its efficacy in addressing the unique challenges of time-series modeling (see Table \ref{tab:performance_comparison}). 

\begin{color}{\differencecolor}
\subsection{On the Role of ICA in Gait Decoding}

Our control experiment with ICA-based artifact removal revealed a trade-off between artifact suppression and neural information preservation. While ICA effectively reduced motion- and eye-related artifacts, performance decreases in stronger models such as EEG2GAIT suggest that task-relevant neural components were inadvertently removed. On average, 8--10 components (over 15\% of total) were discarded, which partly explains the diminished effectiveness of ICA for high-performing models. 

Since ICA requires offline decomposition and manual or semi-automatic component selection, it is not directly compatible with real-time BCI systems. Our results indicate that EEG2GAIT achieves high decoding accuracy under online-compatible preprocessing (i.e., band-pass filtering) without relying on ICA. This suggests that retaining neural information—even at the cost of some residual artifacts—benefits gait decoding more than aggressive artifact removal. Consequently, we advocate band-pass filtering alone for real-time applications, balancing artifact suppression and information retention.

\end{color}

\subsection{Saliency Map Insights into Cortical Dynamics}

Feature importance analysis using saliency maps revealed a consistent focus on sagittal-line electrodes in the frontocentral region, particularly around Cz, FC1, FC2, Fz, \textcolor{\differencecolor}{CPz}, CP1, and CP2. These electrodes correspond to the bilateral leg motor area, aligning with prior findings on the neural representation of lower-limb motor control~\cite{zhang2017multiple, wei2023decoding}. The consistency of these results across participants underscores the neurophysiological relevance and generalizability of the extracted features, confirming the model’s robustness against individual variations in head physiology.
\begin{color}{\differencecolor}
\subsection{Practical Applications in Assistive Technologies}

Beyond benchmarking performance, the proposed EEG2GAIT framework holds considerable potential for practical applications in assistive technologies. First, our preliminary experiments suggest that variations between EEG-predicted gait dynamics and actual recorded trajectories are associated with the rehabilitation stages of stroke patients. This suggests that EEG2GAIT could serve as a fine-grained tool for evaluating motor recovery, thereby enabling objective assessment of rehabilitation progress and providing clinicians with a quantitative biomarker for monitoring patient-specific recovery trajectories.  

Second, EEG2GAIT’s capability to predict lower-limb joint angles with high temporal resolution enables precise identification of gait phases, particularly the swing phase. Accurate detection of these critical gait events is especially important in rehabilitation, as it allows clinicians and assistive devices to pinpoint problematic intervals within the gait cycle. Importantly, this temporal precision permits targeted neural modulation during specific phases of dynamic movement, supporting phase-specific interventions that can enhance motor recovery. Additionally, it facilitates real-time control of lower-limb exoskeletons and prosthetic devices, where timely recognition of motor intent improves synchronization between the user and the device, thereby enhancing both safety and functional effectiveness in brain-controlled assistive systems.

\subsection{Limitations}

Despite its promising performance, EEG2GAIT has two main limitations. First, most practical applications of brain-controlled exoskeletons are based on motor imagery, whereas our current model relies on decoding EEG signals recorded during actual walking. This gap may limit direct applicability in scenarios where overt movement is not possible.  

Second, EEG2GAIT is currently a subject-dependent model, which means that its generalization to unseen individuals is limited. Future work will explore subject-independent or transfer learning strategies to improve cross-subject generalizability and reduce calibration requirements.  

\subsection{Future Work}

To address the identified limitations and advance the clinical applicability of EEG2GAIT, several research directions will be pursued in future work.

\textbf{Cross-Subject Generalization:} Our analysis of saliency maps has revealed important insights that will guide the development of subject-independent models. While we observed consistent patterns across most subjects, where brain regions associated with lower-limb movement consistently showed higher saliency than other areas, each subject still exhibits distinct saliency map characteristics. Notably, we discovered that subjects' saliency maps can be clustered, indicating that some subjects share higher similarity in their neural activation patterns than others.

Based on these findings, \textbf{saliency map-based transfer learning} between subjects represents a promising optimization direction for developing generalized models. We plan to develop a hierarchical transfer learning framework that: (1) identifies subject clusters based on saliency map similarity using unsupervised learning algorithms, (2) applies cluster-specific fine-tuning strategies to new subjects, and (3) incorporates few-shot learning techniques to minimize calibration data requirements. This approach could significantly reduce the calibration time required for new users while maintaining prediction accuracy.

\textbf{Motor Imagery Integration:} Future work will explore bridging the gap between actual walking and motor imagery by investigating shared neural representations between these two modalities. We will develop domain adaptation techniques that can transfer knowledge learned from actual walking data to motor imagery scenarios, enabling the system to work with patients who cannot perform overt movements.

\textbf{Clinical Translation:} From a clinical perspective, we will focus on validating the model with patients having different neurological conditions (stroke, spinal cord injury, etc.) and developing adaptive algorithms that can adjust to disease progression or recovery. Additionally, we will integrate EEG2GAIT with existing rehabilitation protocols and assistive devices to demonstrate its practical utility in real-world therapeutic settings.

\end{color}
\section{Conclusion}

In this work, we proposed a hierarchical graph-based model, EEG2GAIT, for decoding gait dynamics from EEG signals. The model integrates a Hierarchical GCN Pyramid for spatial feature extraction with a Hybrid Temporal-Spectral Reward (HTSR) Loss to capture intricate temporal and spectral features. 
We also contributed a new Gait-EEG dataset. Experiments on the GED dataset and the publicly available MoBI dataset demonstrated that EEG2GAIT surpasses state-of-the-art methods in accuracy and robustness for joint angle prediction.
Ablation studies validated the contributions of the HGP module and HTSR Loss, while saliency map analyses highlighted the spatial correlates of motor decoding, underscoring the importance of central motor regions. These results demonstrate the potential of EEG-based gait decoding for advancing brain-computer interface (BCI) applications in neurorehabilitation and assistive technologies.

Future studies could extend the dataset to diverse populations and explore adaptive graph structures and attention mechanisms to improve decoding of complex neural patterns.

\printbibliography
\end{document}